\documentclass[longnamesfirst,apjl]{emulateapj}
\usepackage{apjfonts}

\newcommand{\beq}{\begin{equation}}
\newcommand{\eeq}{\end{equation}}
\newcommand{\bea}{\begin{eqnarray}}
\newcommand{\eea}{\end{eqnarray}}
\begin{document}

\title{The Cluster-Merger Shock in 1E 0657$-$56: Faster than the Speeding Bullet?}
\author{
Milo\v s Milosavljevi\'c\altaffilmark{1},
Jun Koda\altaffilmark{1}
Daisuke Nagai\altaffilmark{2}, 
Ehud Nakar\altaffilmark{2}, and
Paul R. Shapiro\altaffilmark{1}
}
\altaffiltext{1}{Department of Astronomy, University of Texas, 1 University Station C1400, Austin, TX 78712.}
\altaffiltext{2}{Theoretical Astrophysics, Mail Code 130-33, California Institute of Technology, 1200 East California Boulevard, Pasadena, CA 91125.  }

\righthead{HOW FAST IS THE BULLET CLUSTER?}
\lefthead{MILOSAVLJEVI\'C ET AL.}

\begin{abstract}
Shock waves driven in the intergalactic medium during the merging of galaxy clusters have been observed in X-ray imaging and spectroscopy.  Fluid motions inferred from the shock strength and morphology can be compared to the cold dark matter (CDM) distribution inferred from gravitational lensing. A detailed reconstruction of the CDM kinematics, however, must take into account the nontrivial response of the fluid intracluster medium to the collisionless CDM motions.  We have carried out two-dimensional simulations of gas dynamics in cluster collisions. We analyze the relative motion of the clusters, the bow shock wave, and the contact discontinuity and relate these to X-ray data. We focus on the ``bullet cluster,'' 1E 0657$-$56, a near head-on collision of unequal-mass clusters, for which the gas density and temperature jumps across the prominent bow shock imply a high shock velocity $4,700\textrm{ km s}^{-1}$. The velocity of the fluid shock has been widely interpreted as the relative velocity of the CDM components. This need not be the case, however.  An illustrative simulation finds that the present relative velocity of the CDM halos is $\sim 16\%$ lower than that of the shock. While this conclusion is sensitive to the detailed initial mass and gas density profile of the colliding clusters, such a decrease of the inferred halo relative velocity would increase the likelihood of finding 1E 0657$-$56 in a $\Lambda$CDM universe.


\end{abstract}

\section{Introduction}
\label{sec:intro}

\setcounter{footnote}{0}

During the growth of structure in the universe, nonlinear cold dark
matter (CDM) halos collide and merge hierarchically.  
When halos collide, the dark matter of one can, if collisionless, 
pass through that of the other, unlike the fluid baryonic gas in the halos.
Since the collisions are frequently supersonic for this gas, 
it passes through shocks where it heats and virializes in the final halo.    Currently, an X-ray surface
brightness and temperature discontinuity identifiable as a merger shock with Mach number significantly above
unity has been detected with {\it Chandra} in two galaxy
clusters: 1E 0657$-$56 \citep{Markevitch:02,Markevitch:06} and A520 \citep{Markevitch:05}.
In both cases, X-ray maps exhibit a bow-shock like temperature and
density jump.
For 1E 0657$-$56, the spatial segregation of the X-ray emitting plasma from the peaks of the mass distribution detected with gravitational lensing has been interpreted as the first direct proof of the existence of dark matter \citep{Clowe:06a}.

The strength and geometry of these shock fronts, and the structure of the contact discontinuity, depend on the  kinetic energy and the detailed
gravitational and gas density profiles of the merging components. 
If one could recover the collisionless CDM dynamics from the fluid
dynamics of the X-ray emitting gas, then the kinematics of merger shocks would be a powerful probe of the nature and clustering of CDM (e.g., \citealt{Markevitch:05,Clowe:06a}).  
However, the two
components are coupled only by gravity and behave differently in the
merging process. 
Recently, it has been assumed that the gas shock velocity inferred from the X-ray measurements of 1E 0657$-$56 equals that of the colliding clusters, implying a rare merger event that might be in conflict with the statistical expectations of the CDM model.
We critically examine this assumption by simulating gas dynamics in cluster mergers and 
studying the
response of the intergalactic medium (IGM).

In the cluster-merger in 1E 0657$-$56\footnote{With an
average X-ray temperature of $14.1\pm0.2\textrm{ keV}$
\citep{Markevitch:02,Markevitch:06,Andersson:06,Markevitch:07}, the cluster 1E
0657$-$56 at redshift $z=0.296$ \citep{Tucker:98} is the hottest and
X-ray brightest known cluster.}, transverse motion of the merger shock ($\approx 4,700\textrm{ km s}^{-1}$; \citealt{Markevitch:06}) greatly exceeds relative radial motion of galaxies in the two components ($\approx 600\textrm{ km s}^{-1}$; \citealt{Barrena:02}). 
This suggests that the
shock was driven in the IGM of the larger component
by a near head-on supersonic passage of the smaller component. Inside
the bow shock, another bow-shaped discontinuity with a reverse
temperature jump (lower temperature on the convex side) is seen; this
has been interpreted as the contact discontinuity separating the
shocked and ram-pressure stripped 
IGM of the two merging components. 
\citet{Takizawa:05,Takizawa:06} recovered the three basic features (a bow shock, a contact discontinuity, and ram-pressure stripping) in hydrodynamical simulations of cluster collisions, but did not address the relative kinematics of the CDM
and the fluid component.

We study the formation and
propagation of the merger shock wave and the evolution of the contact
discontinuity, and relate these observable features
of X-ray surface brightness and temperature maps to CDM kinematics. In
\S~\ref{sec:simulations}, we describe the numerical method,
report the general features
of the simulations, and  analyze a
particular run that approximates the observed 1E 0657$-$56 (the conclusions also apply to other cluster mergers).  In \S~\ref{sec:discussion}, we discuss implications for CDM clustering.   The standard cosmological parameters consistent with WMAP \citep{Spergel:06} are assumed throughout.

\section{Simulations of Cluster Merger Shocks}
\label{sec:simulations}

\subsection{The Algorithm and Initial Conditions}
\label{sec:initial}

The simulations were carried out with the thoroughly-tested Eulerian code ASC
 FLASH \citep{Fryxell:00} in two spatial dimensions. Its adaptive mesh
 refinement capability allowed us to simulate a large spatial
 domain with dimensions of $20\textrm{ Mpc}$ and simultaneously
 achieve a spatial resolution of $\sim 1\textrm{ kpc}$ at 
 fluid discontinuities. 
The simulations were carried out in cylindrical ($r$,$z$) coordinates
in an inertial frame.  Axial symmetry restricts the
simulations to head-on cluster collisions.  To avoid artifacts which can result when a perfectly head-on collision occurs with cuspy halo profiles and to approximate the
conditions in non--head-on collisions, the
central dark matter and gas density of the larger cluster was artificially reduced.  The
gravitational potential of each of the clusters is spherically
symmetric and a fixed function of distance from the center.  The clusters are placed at separation $D_{12}$
and are allowed to move under each other's gravity.  The acceleration
of each cluster is evaluated at its center.  The effect of the Hubble flow on cluster motion is ignored because it only
affects the early infall, while we are interested in the dynamics 
of the two halos and their gas content at pericenter passage.  We
also ignore the dynamical friction drag that reduces
the halo velocity during pericenter passage. The fluid flow and its relation to CDM dynamics
within a few hundred kiloparsecs from the center of the smaller halo should not
be different in the presence of dynamical friction, because the
gravitational field of the dynamical wake that trails the smaller
CDM halo will not have strong gradients (tides) on such scales.  Finally, we ignore the tidal stripping of the smaller CDM halo which could be substantial on larger scales than are relevant here.

For the mass distribution associated
with the gravitational potential we consider the \citet{Navarro:97} profile
(``NFW'') $\rho_{\rm NFW}=\rho_0 (r/r_{\rm s})^{-1} (1+r/r_{\rm
s})^{-2}$, and the ``core'' profile $\rho_{\rm core}=\rho_0
(1+r/r_{\rm s})^{-3}$.  The profiles were normalized such that
the mass enclosed within a fixed radius $r_{500}$ equals the
mass of the cluster $M_{500}$.   The
gas density profile was allowed to differ from that of the CDM
profile. At the start of each simulation,
monatomic gas with adiabatic polytropic equation of state
($\gamma=5/3$) is set in hydrostatic equilibrium within each halo.
Explicit viscosity, heat conduction, cooling, and
self-gravity of the gas are ignored. Runs were carried out for a  
variety of merger velocities and density
profiles that included NFW and core
CDM profiles, as well as cuspy (divergent) and non-cuspy gas
density profiles.  In divergent CDM density profiles,
a small
constant-density core ($20\textrm{ kpc}$) in hydrostatic equilibrium was applied 
in the center to resolve the 
fluid gradients with multiple resolution elements.
 The cluster gas fraction was $f_{\rm gas}\Omega_{\rm b}/\Omega_{\rm m}$, where $\Omega_{\rm b}/\Omega_{\rm m}\approx 0.16$ is the
cosmological baryon abundance and $f_{\rm gas}\approx 0.7$ \citep{Afshordi:06}.\footnote{All gas-dynamical results are independent of this assumed value of $f_{\rm gas}$, except that the X-ray surface brightness $\propto f_{\rm gas}^2$.}   The mean molecular weight was $\mu=0.59$, corresponding to $75\%$ hydrogen and $25\%$ helium by mass, fully ionized. 
Numerical convergence
was ascertained by resimulation at increased spatial resolution.

\subsection{General Features of Cluster Merger Shocks}
\label{sec:general}

\begin{figure}
\plotone{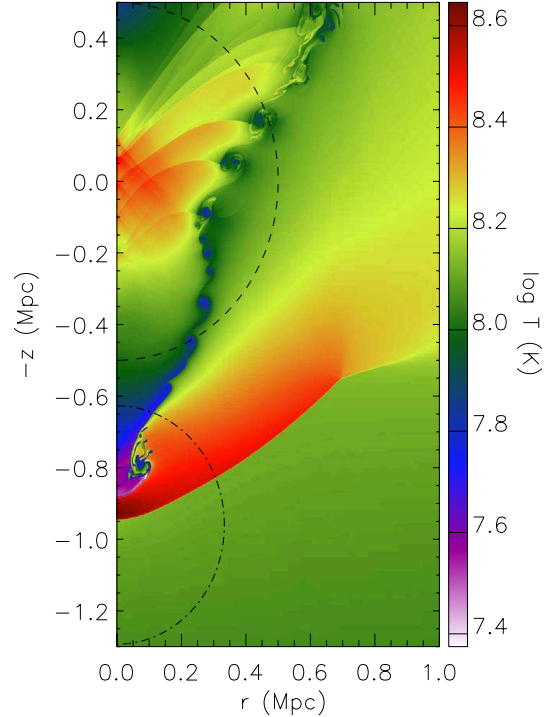}
\caption{
Temperature 
in the meridional plane after cluster pericenter passage. The larger CDM halo is located at
$(r,z)=(0,0)$.  The circles indicate the scale
radii $r_{\rm s}$ of the two halos.\label{fig:denstemp}}
\end{figure}

Two prominent features seen in all simulations are a bow shock
wave, and a contact discontinuity (cold front).
The formation of a shock
wave during cluster merger was discussed by
\citet{Markevitch:00}; 
our results are consistent with their picture.  
Opening angle and
the radius of curvature of the nose of the bow shock are
sensitive to the details of the simulation, but both are larger than
those expected in steady-state bow shocks driven by spherical hard
spheres moving with a constant velocity in a uniform medium (e.g.,
\citealt[and references therein]{Farris:94,Vikhlinin:01}). Material on
the convex side of the contact discontinuity originates in the
smaller cluster. During the collision, a reverse shock propagates into
the smaller cluster; a fraction of the smaller cluster traversed by the
reverse shock is heated, but the dense core of the  cluster
remains cold as the reverse shock weakens inside it.  
 The wings of the contact discontinuity are
Kelvin-Helmholtz (KH) unstable (in real clusters, the KH instability will be modified by magnetic stresses).  A nonlinear instability disrupts the
coherence of the surface of the discontinuity and creates a
narrowing in the axial diameter of the discontinuity (a neck) just
behind the nose, giving it a mushroom-like or
bullet-like appearance, as evident in the temperature map in 
Figure \ref{fig:denstemp}.  At later times (not shown), the Rayleigh-Taylor instability eats into and disrupts the bullet from the rear, as predicted by \citet{Markevitch:07}.

\begin{figure}
\plotone{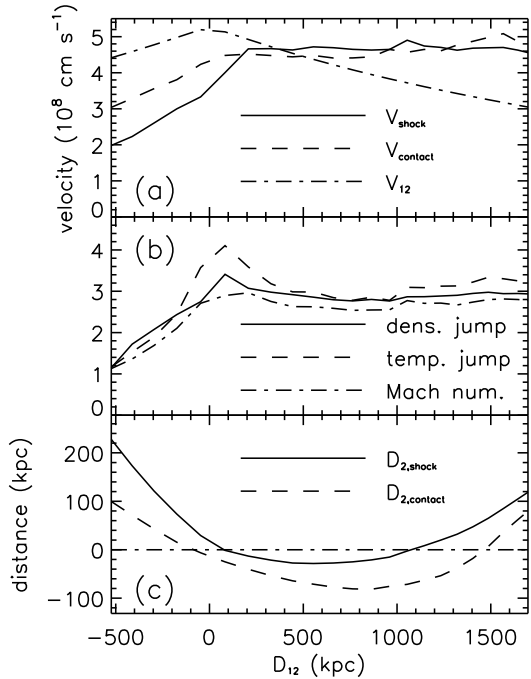}
\caption{
(\emph{a}) Velocity of the shock ({\it solid}) and of the contact discontinuity ({\it dashed}) relative to the pre-shock gas upstream, and relative
velocity of the CDM halos ({\it dot-dashed}), all as 
functions of the time-varying separation between the halos $D_{12}=z_2-z_1$.
(\emph{b})  The density ({\it solid}) and temperature jump ({\it dashed}) at the
shock, and the shock Mach number ({\it dot-dashed}).
(\emph{c}) 
Distance of the shock ({\it solid}) and the contact
discontinuity ({\it dashed}) from the center of the smaller
halo.\label{fig:shock}}
\end{figure}

In most simulations, during 
pericenter passage, the shock and contact discontinuity are slower
than the relative velocity of the two CDM halos (shock velocity is
calculated in the local rest frame on the upstream side of the shock).  This is
due to the ram-pressure force acting on the gas in the center
of the smaller cluster but not on the halo. In many runs some degree of ram-pressure
stripping is evident in the center of the smaller cluster (we call
the smaller cluster ram-pressure stripped when cold gas originating in
the smaller cluster has been expelled from the center of the smaller
halo). However, later, as the halos climb out of each
other's gravitational potential well and decelerate, but the shock and
the contact discontinuity do not decelerate
appreciably over a longer period (Fig.~\ref{fig:shock}\emph{a}). 
The lack of
deceleration of the shock and the cold front was noticed,
though not quantified, in previous simulations of cluster mergers
 \citep[and references therein]{Markevitch:07}. It could result from the drop in ram pressure as the cold
bullet propagates into a thinning cluster atmosphere and from the gravitational tide of the halo.  
After the cold
gas originally in the smaller cluster has been ram-pressure stripped, the
stripped gas lags behind the smaller  halo, but eventually its velocity becomes
larger than that of the halo, which implies that the gas can later catch up with the halo again.

Before proceeding to discuss a specific simulation approximating the observed bullet cluster, we strongly emphasize that the shock kinematics and morphology are \emph{extraordinarily sensitive} to the parameters of the simulation.  In particular, a small variation ($\sim 25\%$) in the cluster mass ratio or density profile can lead to vastly different degrees of ram-pressure stripping, or the opening angle of the bow shock.

\subsection{A Model of 1E 0657$-$56}
\label{sec:application}

General results discussed in \S~\ref{sec:general} are illustrated by
a specific run that came the closest to reproducing the observed
properties of 1E 0657$-$56.  Cluster
masses in the run were $M_{500,1}=1.27\times10^{15}M_\odot$ and
$M_{500,2}=2.54\times10^{14}M_\odot$,  cluster radii were
$r_{500,1}=1.5\textrm{ Mpc}$ and $M_{500,2}=1\textrm{ Mpc}$, and the
cluster scale radii were $r_{\rm s,1}=500\textrm{ kpc}$ and $r_{\rm
s,2}=333\textrm{ kpc}$.\footnote{The mass ratio of the two halos in the simulation, $M_{500,1}/M_{500,2}=5$,
is about a half of the ratio $\sim 10$ inferred from lensing data 
\citep{Clowe:04,Clowe:06a,Bradac:06}. Some fraction of the smaller cluster's 
initial mass will have been tidally stripped during the merger, possibly reconciling 
our initial conditions with lensing.  
The total luminosity $L_{\rm tot}$ of
galaxies in each cluster can be used to estimate the cluster 
mass via the relation $L_{\rm tot}\propto M_{500}^\beta$  \citep{Lin:04,Miller:05,Cooray:05}, 
where $\beta\lesssim 1$ is a
bandpass-dependent factor.  For the
$R$-band we adopt $\beta_R=0.65$.  \citet{Barrena:02} report $L_{{\rm
tot},1,R}=10^{12}L_\odot$ and $L_{{\rm
tot},2,R}=0.2\times10^{12}L_\odot$, implying a
mass ratio of $M_{500,1}/M_{500,2}\approx 12$, in agreement with lensing.
The difference of the mass ratios may also be attributed to
a slightly non--head-on collision in 1E 0657$-$56.
Indeed, the lensing maps are
consistent with a northerly passage of the smaller cluster
inducing the observed SW$-$NE tidal distortion in the larger cluster.
Then, the lower density column met by the small
cluster would result in equivalent shock dynamics if the
smaller cluster were smaller than in the simulation.}
The density profile of the larger
cluster was the non-divergent $\rho_{\rm core}$ with central density $\approx 2.3\times10^{-3}M_\odot\textrm{ pc}^{-3}$ (again, to approximate
the conditions in a non--head-on merger with pericenter passage at a distance $\sim r_{s,1}$), and, in the smaller cluster, it
was the NFW profile $\rho_{\rm NFW}$.  The gas density in both
clusters was proportional to the CDM density, $\rho_{\rm
gas}=(1+\Omega_{\rm m}/f_{\rm gas}\Omega_{\rm b})^{-1} \rho_{\rm CDM}$. The halos initiated infall from relative rest at separation $D_{12}=4.6\textrm{ Mpc}$ and their maximum relative velocity at pericenter was $V_{\rm max}\approx5,270\textrm{ km s}^{-1}$.\footnote{All dimensionless
numbers (such as the shock Mach number, and the gas-to-dark matter
velocity ratio) and distances (such as the radius of curvature of the
shock) are invariant under a uniform scaling of the CDM 
density that preserves lengths.  The scaling preserves 
 the mass to temperature ratio.}

\begin{figure}
\plotone{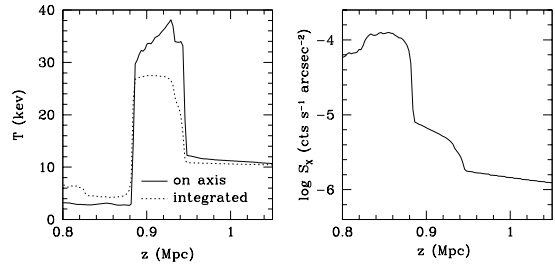}
\caption{Temperature profile on the central axis ({\it left panel, solid line}); emissivity-weighted integrated temperature profile ({\it left panel, dotted line}); and $0.8-4\textrm{ keV}$ surface brightness on the central axis ({\it right panel}). The center of the larger CDM halo is located at $z=0$. The bow shock at $z\sim 0.94\textrm{ Mpc}$ is split into a main shock and a weaker subshock.\label{fig:cut}}
\end{figure}

Figure \ref{fig:denstemp} shows a snapshot of the meridional
temperature distribution when the distance of the shock to
the larger CDM halo is about $1\textrm{ Mpc}$. The larger and smaller
cluster are moving vertically upward and downward in the figure,
respectively. The cold gas bullet is $\sim 100\textrm{ kpc}$ in radius, excluding visible fingers of cold gas embedded in the hot post-shock
medium.  The center of the smaller halo is located ahead
(below) the contact discontinuity, and almost exactly at the location
of the bow shock, as found in X-ray and lensing maps of 1E 0657$-$56 \citep{Clowe:04,Clowe:06a,Bradac:06}. This is also evident in Figure \ref{fig:shock}{\emph c}
showing the position of the bow shock and the contact discontinuity
relative to the position of the smaller halo. Ram-pressure stripping
at the center of the smaller cluster occurs at $D_{12}=0$, but the bow
shock and the contact discontinuity catch up with the halo at
$D_{12}\approx 1.2\textrm{ Mpc}$ and $\approx 1.5\textrm{ Mpc}$
respectively.  

\begin{figure}
\plotone{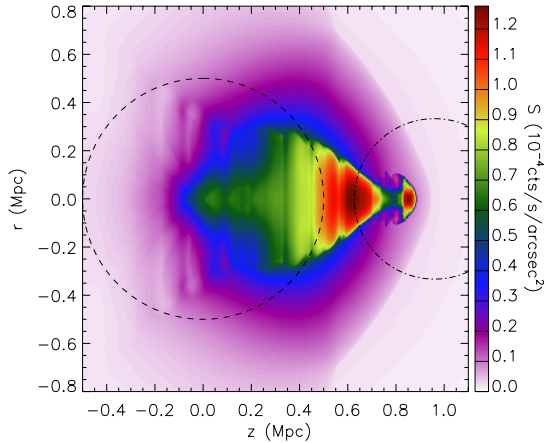}
\caption{Surface brightness map of the simulated bullet cluster in the $0.8-4\textrm{ keV}$ band, assuming a frequency-independent effective area of $400\textrm{ cm}^2$.\label{fig:surface}}
\end{figure}

The pre- and post-shock axial temperatures are $12\textrm{ keV}$ and $34\textrm{ keV}$, respectively (see Figure \ref{fig:cut}). Figure \ref{fig:shock}{\emph b}
shows the evolution of the density and temperature jumps across the
shock, and its Mach number, as a function of the 
halo separation. The Mach number is $\approx3$ when the separation
equals the  projected separation of lensing density peaks $D_{12}\approx 720\pm25\textrm{ kpc}$ \citep{Clowe:04,Clowe:06a,Bradac:06}. The temperatures and shock strength are compatible with the results of \citet{Markevitch:06}, who, for a $\gamma=5/3$ polytropic equation of state and instant electron-ion equilibration, derived the Mach number from deprojected density and temperature jumps in the bow shock in 1E 0657$-$56.  
The separation of the shock and the contact
discontinuity is $\sim 70\textrm{ kpc}$, somewhat smaller than in the
observed system.
Figure \ref{fig:surface} shows the predicted X-ray surface brightness map for the simulated cluster.

Figure \ref{fig:shock}{\emph a} shows the evolution of the velocities
of the shock, the contact discontinuity, and the relative velocity of
the two CDM halos. While the cluster velocity decreases as the small
halo climbs the potential of the larger halo, the shock velocity
remains constant.  The lack of deceleration of the shock can be explained by a
combination of factors; in addition to gravitational forces, the negative density and
temperature gradients into which the shock propagates tend to
increase its strength.  At $D_{12}=720\textrm{ kpc}$, the
velocity of the shock is  $V_{\rm sh}\approx 4,800\textrm{ km s}^{-1}$, consistent with $V_{\rm sh}\sim 4,740^{+710}_{-550}\textrm{ km s}^{-1}$ inferred for 1E 0657$-$56 by Markevitch (as cited in \citealt{Farrar:06}, private communication). However, the
relative velocity of the halos in our simulation is much less, $V_{\rm sub}\sim 4,050\textrm{ km s}^{-1}$.\footnote{If the true halo separation is larger than the projected separation $720\textrm{ kpc}$, the predicted subcluster velocity $V_{\rm sub}$ will be $< 4,050\textrm{ km s}^{-1}$.}

\section{Discussion and Conclusions}
\label{sec:discussion}

Recent analyses of 1E 0657$-$56 (e.g.,
\citealt{Hayashi:06,Clowe:06a,Bradac:06,Farrar:06}) assume that the
shock velocity  equals the instantaneous
relative velocity of the CDM halos, and notice that it is much larger
than the virial velocity of the larger cluster.
\citet[hereafter HW06]{Hayashi:06} and \citet[hereafter FR06]{Farrar:06} estimate the probability that a halo in the $\Lambda$CDM universe has a subhalo with velocity larger than that
of the bow shock in 1E 0657$-$56.  We assess a correction that should be applied to the inferred probabilities to account for the relative motion of the shock and the subhalo.
HW06 
calculate the likelihood of finding a halo-subhalo pair with such relative
velocity by extrapolating results of the Millenium cosmological $N$-body simulation. They 
find that it decreases rapidly with $V_{\rm sub}$,
$\log[N_1(>V_{\rm sub})/N_{\rm hosts}] = -(V_{\rm
sub}/v_{10}V_{200})^\alpha$, where $v_{10}\approx1.55$ and
$\alpha\approx 3.3$.
The probability depends on the assumed values of $V_{\rm sub}$ and $V_{200}$, in which HW06 and FR06 differ.\footnote{We do not test the accuracy of the cited values of $V_{200}$, but the lensing data seem to indicate that HW06's value of $V_{200}$ is an overestimate.}
HW06 adopt a shock velocity of $V_{\rm sh}\approx 4,500_{-800}^{+1100}\textrm{ km s}^{-1}$ \citep{Markevitch:04} as the relative velocity $V_{\rm sub}$ of the two halos and adopt $V_{200}\approx 2,380\textrm{ km s}^{-1}$ for the virial velocity.  With these, they infer that the probability of most
massive subhalo having such a velocity is $N_1(>V_{\rm sub})\approx 1\%$, whereas when they adopt a lower shock velocity consistent with
uncertainties of $V_{\rm sub}=3,700\textrm{ km s}^{-1}$, the
probability is $N_1(>V_{\rm sub})\approx 10\%$.
In our simulation
(\S~\ref{sec:application}), we find that, when the halos are
separated by the observed projected distance $D_{12}=720\textrm{ kpc}$, the velocity of the shock and that of the subhalo differ by $(V_{\rm sh}-V_{\rm sub})/V_{\rm sub}\approx 16\%$.  Thus, the probabilities reported by HW06 should be corrected and should read
$\approx 8\%$ and $\approx 27\%$,
respectively.
FR06 adopt a somewhat higher shock velocity $V_{\rm sh}\approx 4,740\textrm{ km s}^{-1}$ and significantly lower virial velocity $V_{\rm 200}\approx 1,740\textrm{ km s}^{-1}$. These values imply a very low probability of $\approx3.7\times10^{-7}$.\footnote{HW06 and FR06 also report $\times5$ lower probabilities after counting only post-collision systems.}  The probability is significantly higher but remains very low, $\approx 2.4\times10^{-4}$, when the $16\%$ correction for the difference between the shock velocity and CDM halo velocity is taken into account.

In conclusion, our 2D hydrodynamical simulations of merging galaxy clusters show that, in a case like the ``bullet cluster'' 1E 0657$-$56, the halo collision velocity need not be the same as the intergalactic gas shock velocity.
While the kinematics of the shock is sensitive to the details of the cluster structure, the instantaneous shock velocity can exceed the relative velocity of CDM halos by at least $\sim 16\%$. Any attempt to relate shock kinematics to the details of hierarchical clustering and halo assembly must take into account the nontrivial response of the IGM.  Published estimates of the likelihood of finding a configuration resembling 1E 0657$-$56 in a $\Lambda$CDM universe may require upward revision by a factor of $2$ to $700$.

\acknowledgements

The research was supported in part by the Sherman Fairchild
Foundation and NASA
Astrophysical Theory Program grant NNG04G177G.
The software used in this work was developed in part by
the DOE-supported ASC / Alliance Center for Astrophysical
Thermonuclear Flashes at the University of Chicago.

\end{document}